\documentclass[a4paper,12pt,fleqn]{article}
\usepackage[dvipdfmx]{graphicx}
\usepackage{mediabb}
\usepackage{url}
\usepackage[dvipdfmx]{graphicx}
\usepackage{amsmath}
\usepackage{multirow}
\usepackage[latin1]{inputenc}             
\usepackage{bm}
\usepackage{setspace}
\usepackage{subfigure}
\usepackage{color}
\usepackage{caption}
\usepackage{amssymb}
\usepackage{cite}

\begin{document}

\title{Local empirical Bayes correction for Bayesian modeling
}

\author{Yoshiko Hayashi\footnote{yoshiko-hayashi@omu.ac.jp\\
This paper is a modified version of my paper published in Osaka Keidai Ronshu, vol.68, no.4, pp.161-172, 2017.  (in Japanese: $\rm{https://doi.org/10.24644/keidaironshu.68.4\_161}$) }\\
Osaka Metropolitan University\\
}

\maketitle

\begin{abstract}

The James-Stein estimator has attracted much interest as a shrinkage estimator that yields better estimates than the maximum likelihood estimator. 
The James-Stein estimator is also very useful as an argument in favor of empirical Bayesian methods. However, for problems involving large-scale data, such as differential gene expression data, the distribution is considered a mixture distribution with different means that cannot be considered sufficiently close. Therefore, it is not appropriate to apply the James-Stein estimator. 
Efron (2011) proposed a local empirical Bayes correction that attempted to correct a selection bias for large-scale data.

On the other hand, when constructing Bayesian models for that data, the setting of the prior and hyper-prior distribution significantly impacts the results. In this paper, we develop a hierarchical Bayesian model for large-scale data without assuming a hyper-prior distribution using Efron(2011)'s bias collection method. The results showed that our model works quite well.
\end{abstract}


\newpage

\section{Introduction}

Efron (2011) proposed a local correction method for bias by applying an empirical Bayesian method using Tweedie's formula\footnote{Efron (2011) refers to this correction as Tweedie's formula, based on Robbins' (1956) citation of Tweedie's 1947 paper, but Koenker and Gu (2016) point out that this Bayesian correction was already introduced by Dyson in 1926.}.

The local empirical Bayes correction by Efron (2011) is based on the empirical Bayes method, which uses estimates of the marginal distribution for likelihood and does not need to specify the prior distribution.
Furthermore, the model does not require the normal distribution of the marginal distribution, which is commonly used in the estimation stage when we adopt a kernel density function or a mixture distribution. 
Efron (2011) uses Lindsey's method to estimate the density function of the marginal distribution without assuming a normal distribution for the marginal distribution.
There have been several developments in the estimation of the density function of the marginal distribution since then. For example, Wager (2014) used a nonparametric estimation method, and Simon and Simon (2013) proposed a method for estimating the density function of the marginal distribution based on a higher-order approximation.
 While Tan et al. (2015) proposed a method for estimating the density function of the marginal distribution based on the correlation between data, which is an extension of the method of Simon and Simon (2013).
On the other hand, many Bayesian models are also used in research involving large-scale data, and Bayesian FDR has been developed for Bayesian models to apply the analysis of False Discovery Rate, which is an important analytical method for analyzing large-scale data (M\"{u}ller et al. (2007)). Furthermore, it is useful that we adopt Student-$t$ modeling to be robust against outliers using Bayesian modeling (e.g. Gottardo et al. (2006)). 

Therefore, it is very important to apply local empirical Bayesian corrections to hierarchical Bayesian models.

This paper examines the applicability of local empirical Bayesian correction based on the posterior predictive probabilities obtained from Bayesian models. The rest of the paper is organized as follows. In Section 2, we provide an overview of empirical Bayesian methods, followed by an introduction to empirical Bayes correction by Efron (2011). Section 3 presents empirical Bayesian corrections using posterior probabilities obtained from the estimated Bayesian model, and  Chapter 4 demonstrates the results based on simulation data. Finally, in Chapter 5, we summarize the results and findings.

\subsection{Local Empirical Bayes Correction}


Efron (2011) applied an empirical Bayesian method using the estimated marginal distribution to the local Bayesian correction based on Tweedie's distribution.
In this section, we will introduce the empirical Bayesian correction developed in Efron (2011).

Consider a model in which the likelihood follows Tweedie's distribution.

\begin{equation}
\left\{
\begin{array}{@{\,}l}
\eta \sim g(\eta),\\
y\mid \eta \sim f_{\eta}(y)=\exp[\eta y - \psi(\eta)]f_0(y),\\
\end{array}
\right.
\end{equation}
where $\psi(\eta)$ represents the cumulant generating function, and $f_0(y)$ represents the density when $\eta=0$.
 Under the model, the posterior distribution of $\eta$ is
\begin{equation}
g(\eta | y)=f_{\eta}(y)g(\eta)/f(y),
\end{equation}
where $f(y)=\int _Y f_{\eta}(y)g(\eta)d\eta$.

Let $\lambda(y)=\log(f(y)/f_0(y))$. Then, the posterior distribution can be arranged as follows: 
\begin{equation}
g(\eta | y)=e^{y \eta - \lambda(y)} g(\eta)e^{- \psi(\eta)}.
\end{equation}
The exponential distribution family can be expressed as,
\begin{equation}
f(x|\theta)=exp[p(\theta)K(x)+S(x)+q(\theta)].
\end{equation}

Thus, the mean and variance are as follows:
\begin{equation}
\left\{
\begin{array}{@{\,}l}
E(x)=-\frac{q'(\theta)}{p'(\theta)},\\
VAR(x)=\frac{1}{p'(\theta)^3}\{p''(\theta)q'(\theta)-q''(\theta)p'(\theta)\}.\\
\end{array}
\right.
\end{equation}
 By substuting $y$ into $\theta$ and $\eta$ into $x$ in the above equation, we have $p(y)=y$ and $q(y)=-\lambda(y)$. 
Therefore, we have the expected value and variance as follows:
\begin{equation}
\left\{
\begin{array}{@{\,}l}
E(\eta | y)=\lambda'(y),\\
VAR(\eta | y)=\lambda''(y).\\
\end{array}
\right.
\end{equation}
Taking the logarithm of the density distribution, $l(y)=\log(f(y))$, $l_0(y)=\log(f_0(y))$, we obtain the posterior distribution of $\eta$ is as follows:
\begin{equation}
\eta | y \sim (l'(y)-l_0'(y), l''(y)-l_0''(y)).
\end{equation}

Assuming the normal distribution ($ y \sim N(\mu, \sigma^2)$, from equation (6) $\mu$ = $\sigma^2 \eta$ (or $\eta=\frac{\mu}{\sigma^2}$).  Thus, $l_0(y)=-\frac{y^2}{2\sigma^2}-\log(\sqrt{2\pi \sigma^2})$. 

By using them, we obtain the following posterior mean.
\begin{equation}
\mu | y \sim (y+\sigma^2l'(y), \sigma^2(1+\sigma^2l''(y)))
\end{equation}
Hence, the local empirical Bayesian correction is derived for the location parameter as follows:
\begin{equation}
E(\mu | y)=y+\sigma^2l'(y).
\end{equation}
In other words, the posterior mean of the mean $\mu$ is an unbiased estimate from the data plus a Bayes correction term. Efron (2011) proposes an empirical Bayes-based bias correction that estimates the correction term $l'(y_i)$ based on the marginal distribution from observed data.
\begin{equation}
\hat{\mu}_i \equiv \hat{E}(\mu_i |y_i)=y_i+\sigma^2\hat{l}'(y_i)
\end{equation}
To obtain an estimate of $l'(y_i)$, Efron (2011) estimates the marginal distribution from a Poisson regression of order $J$, based on Lindsey's Method, which assumes that the frequencies of $y$ divided into $K$ bins follow independent Poisson distributions.

 As noted in Efron (2011), when we assume a second-order spline, this empirical Bayes correction is reduced toward the hyperparameters of the positional population as well as the James-Stein estimator.
When $y_i$ follows a normal distribution with mean $\mu_i$ and variance 1, then regardless of the prior distribution of $\mu_i$, the empirical Bayes correction for $\mu_i$ is given by
\begin{equation}
\hat{\mu}_i =y_i+\hat{l}'(y_i)
\end{equation}

\section{Empirical Bayes Correction to Bayesian Modeling}

We investigate the method that applies the empirical Bayes correction of Efron (2011) to Bayesian modeling. We obtain the correction term from the distribution of the posterior statistical scores of the individual location populations, despite the hyperprior distribution is unknown. In other words, we do not require assuming the form of the hyperprior distribution in this method. We treat the posterior statistical score obtained from the posterior distribution of the individual location populations as $y_i$ in equation (12), and estimate $\hat{l'}(y_i)$ from the distribution of the posterior statistical score.

If the posterior statistical scores for the individual location population do not belong to an exponential distribution family, the empirical Bayes correction of Efron (2011) cannot be applied. However, when analyzing large-scale datasets, such as differential gene expression data, individual scores are often obtained as $t$ value rather than $z$ score. As the $t$ distribution does not belong to the exponential distribution family, the empirical Bayes correction proposed by Efron (2011) cannot be applied.
This means that the logarithm of the density function of the $t$ distribution is not a concave function at its tail, and therefore applying empirical Bayesian correction will further increase the bias unless the observed values are sufficiently close to the parameters.

 For this problem, Efron (2011) proposed to apply the empirical Bayes correction by obtaining the lower probability from the distribution function of the $t$ distribution and then using the inverse function of the standard normal distribution to obtain $z$-values that follow the standard normal distribution.

\begin{equation}
z=\Phi^{-1}(F_{\nu}(t)) \hspace{1.0cm} t \sim t_{\nu}(\delta, \kappa)
\end{equation}
where $\Phi$ denotes the distribution function of the standard normal distribution and $F_{\nu}(t)$ denotes the distribution function of the $t$-distribution with degree of freedom $\nu$ and location parameter $\kappa$.

We apply this transformation to the posterior probability instead of $F_{\nu}(t)$ and obtain scores that follow the standard normal distribution.

\subsection{Robust standard deviation in Bayesian model}

 Ventrucci and Scott(2011) defined the posterior predictive probability for the parameter $\theta$ as the mean of the data belonging to the null group being zero as follows:
\begin{eqnarray}
p_{Bi} \equiv Pr\{\theta_i>0 \mid data \},
\end{eqnarray}
When the subject belongs to the null group, this posterior predictive probability follows a uniform distribution (e.g. Storey (2003)). Thus, $p_{Bi}$ is defined as the lower probability of this posterior predictive probability.

We can estimate $p_{Bi}$ from a sufficiently large $A$ times Markov chain Monte Carlo (MCMC) iteration as follows:
\begin{equation}
p_{Bi} \doteq \frac{1}{A}\sum^A_{i=1} I \{\theta_i^{rep} > 0 \mid data \}
\end{equation}
 where $\beta^{rep}$ represents the MCMC replication.

Hence, the following posterior scores using the inverse function of the standard normal distribution follow the standard normal distribution.
\begin{equation}
z_i^B=\Phi^{-1}(p_{Bi}) \hspace{1.0cm}
\end{equation}

We adopt the posterior predictive probability for the parameter using $S_R$ based on $z_i$ in equation (16). However, we should be careful that when we estimate $p_{Bi}$ using MCMC iterations, $p_{Bi}$ would take 0 and 1, which means $z_i$ takes infinities. Moreover, even if we use a large enough $A$, such as 10,000, the smallest $p_{Bi}=0.0001$ or -3.719. Thus, we can not obtain the $z$ score being less than -3.719.  Thus, we have to take care of those values.

\subsection{Estimation of distribution of posterior scores}
Our procedure to estimate the posterior probability is based on Ventrucci and Scott(2011), but we use $\beta^*$ of $p_{Bi}=0.50$ instead of zero to generate the posterior predictive probability for the parameter $\beta$. That is, the median of the posterior distribution as follows:
\begin{eqnarray}
p_{Bi} \equiv Pr\{\beta_i>\beta^* \mid data \}
\end{eqnarray}
By using the $p_{Bi}$ and the inverse function of the normal distribution, we obtain a variable that follows a normal distribution with the mean being the median ($\beta_i^*$). This satisfies the assumption of Tweedie's formula that the distribution belongs to the exponential family. 

In addition, when a distribution follows a complex shape, standardization cannot be performed using the value of the estimated variance. For this problem, Efron and Zhang (2011) define a robust standard deviation as the two halves of the distance from the 16th to the 84th percentile of the $z_i$. We adopt their method to estimate the standard deviation using the percentiles of the posterior score of $\beta_i$.

The posterior score is therefore the median of the posterior distribution ($\beta_i^*$)  divided by the robust standard deviation $(S_R)$ as follows:
\begin{eqnarray}
\hat{z}_i^*=\frac{\beta_i^*}{S_R} 
\end{eqnarray}
If all data belong to the null group, then the score using the robust standard deviation in equation (18) approximates the statistical scores obtained in equation (16).

\subsection{Simulation Analysis}

In this section, we will investigate the performance of our model based on the simulated data.
Efron (2011) shows that the results of simulations using $e^{-\mu}$ for the prior distribution of $\mu$ have good performance from the comparison of the regret. However, he notes that for extreme values such as the 99.9th percentile, the empirical Bayesian approximation is not very accurate.

In this paper, we set the sample size for each dataset to 30 ($n = 30$) and use 1,000 datasets ($N = 1,000$) for the simulations. For simplicity, equal variances are assumed, and the following linear model is used.
\begin{eqnarray}
 y_{ji} =  \alpha_j+\beta_j D_{i} +u_{ji} \hspace{0.5cm} j=1,\ldots,1000. \quad i=1,\ldots,30.
\end{eqnarray}

where $u_{ji} \sim N (0, 1)$ and $D_i$ is a state variable. The state variable takes $D_i = 0$ for null data, and takes $D_i = 1$ for non-null data.

Among the 1,000 data sets, we create 100 non-null datasets, which we divide into two groups. In the non-null datasets, 15 samples have $D_i=1$.

\subsection{Model}

We adopt a normal distribution for the likelihood, and Jeffrey's priors are used for the location and scale parameters, respectively. A summary of the model is as follows:

\begin{equation}
\left\{
\begin{array}{@{\,}ll}
f(y_{ji}|\mu_{ji}, \sigma_{j})= N(\mu_{ji},\sigma_{j}^2),\\
\alpha_{j},\beta_{j}  \stackrel{D}{\sim} Uniform,\\
\log(\sigma_{j})  \stackrel{D}{\sim} Uniform,\\
\end{array}
\right.
\end{equation}

where $\mu_{ji}=\alpha_{j}+\beta D_{ji}.$ OpenBUGS (v.3.2.3) was used for the estimation. The number of iterations was 11,000, and the first 1,000 have been removed.
As in Efron (2011), Poisson regression is used to estimate the distribution of the posterior distribution of scores from the histogram frequencies using a generalized linear model (glm) in R(3.1.1) based on Lindsey's method.

\subsection{Results}

 Figure 1 shows the histograms using the interval 0.05, and Figure 2 shows the distribution of the posterior statistical scores obtained by second and fifth-order Poisson regression analysis using these histograms. These results show that the results using the robust standard deviation as defined in equation (18) provide a good approximation. Thus, it is reasonable to adopt it to our estimation.

We use six simulation scenarios, as shown in Table 1. Figure 3 shows histograms using an interval of 0.25 with data generated from each simulation model. Figure 4 shows the splines and empirical Bayes correction estimated from a fifth-order Poisson regression using the obtained histograms.

\begin{figure}[ht]
\begin{center}
\subfigure[score of robust SD]{
\includegraphics[width=50mm,height=50mm]{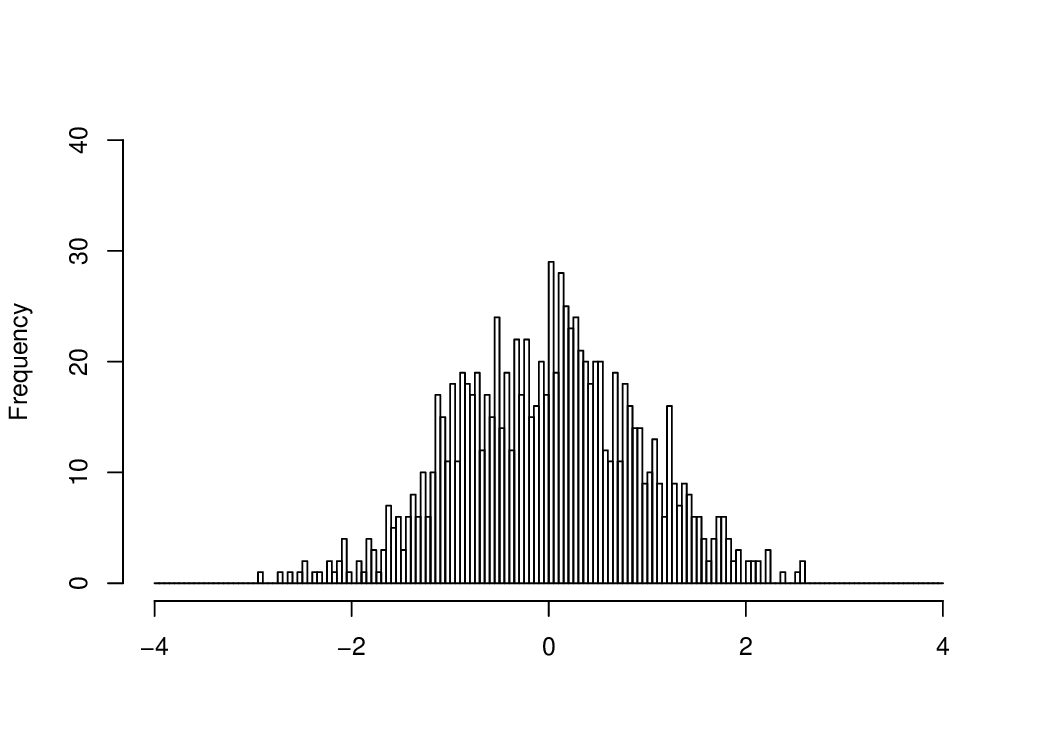}}
\subfigure[Score of inversed function]{
\includegraphics[width=50mm,height=50mm]{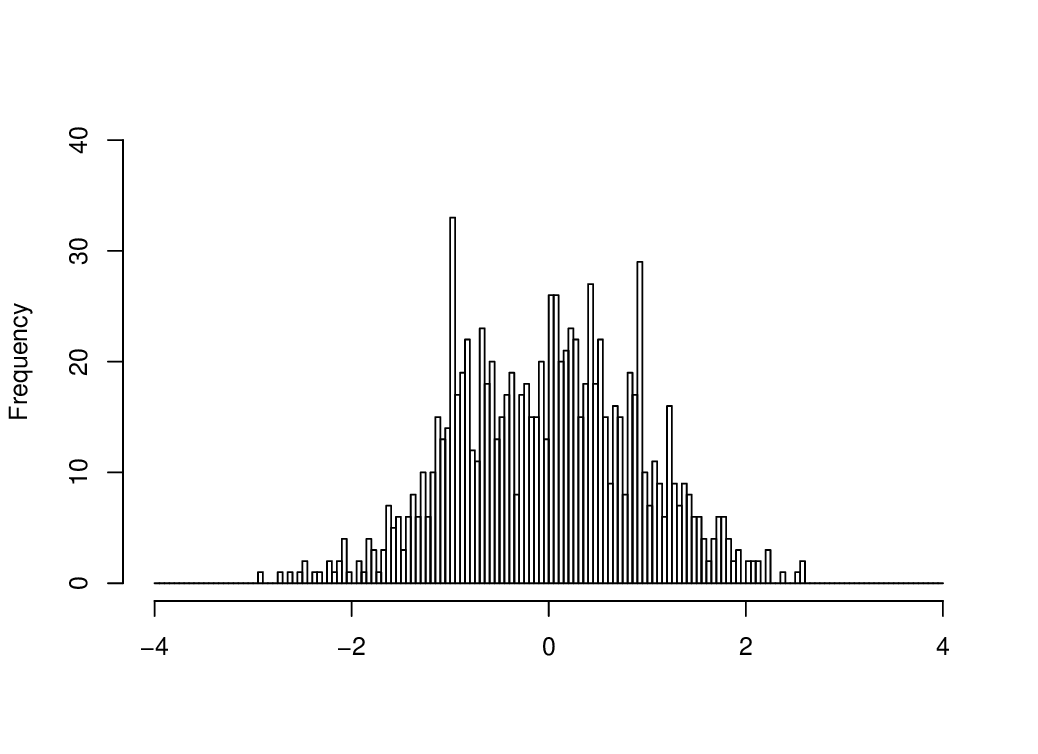}} \\
\caption{Histogram of scores (all null groups)} 
\end{center}
\label{fig:1}
\end{figure}

\begin{figure}[ht]
\begin{center}
\subfigure[2nd order Poisson regression]{
\includegraphics[width=50mm,height=50mm]{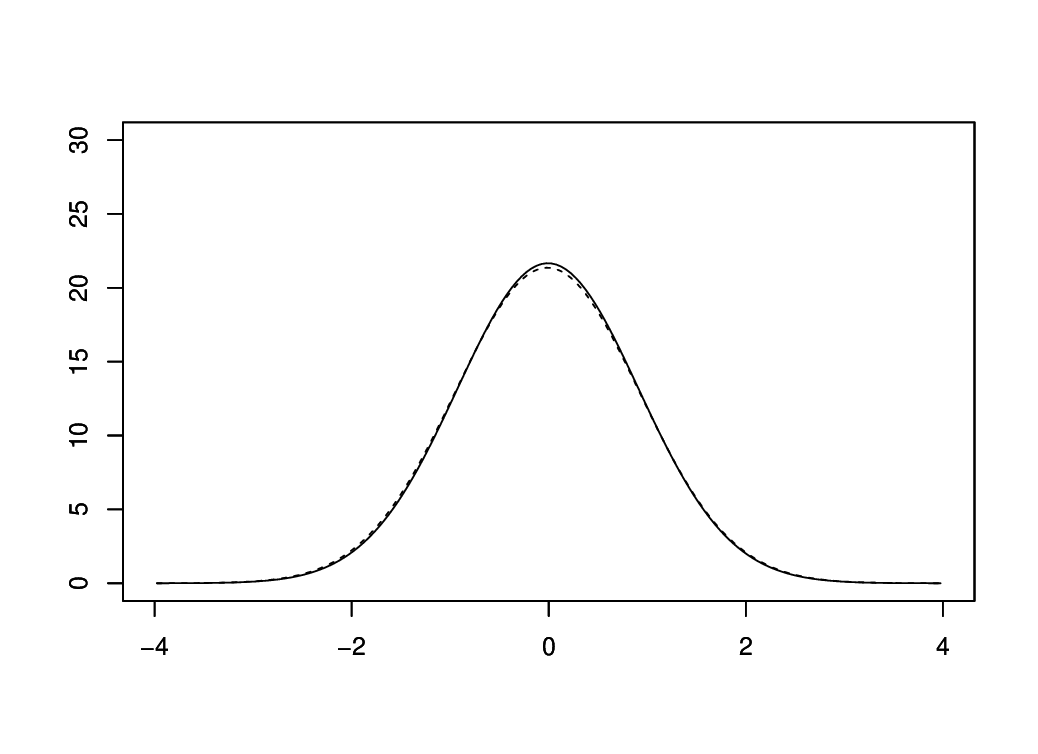}}
\subfigure[5th order Poisson regression ]{
\includegraphics[width=50mm,height=50mm]{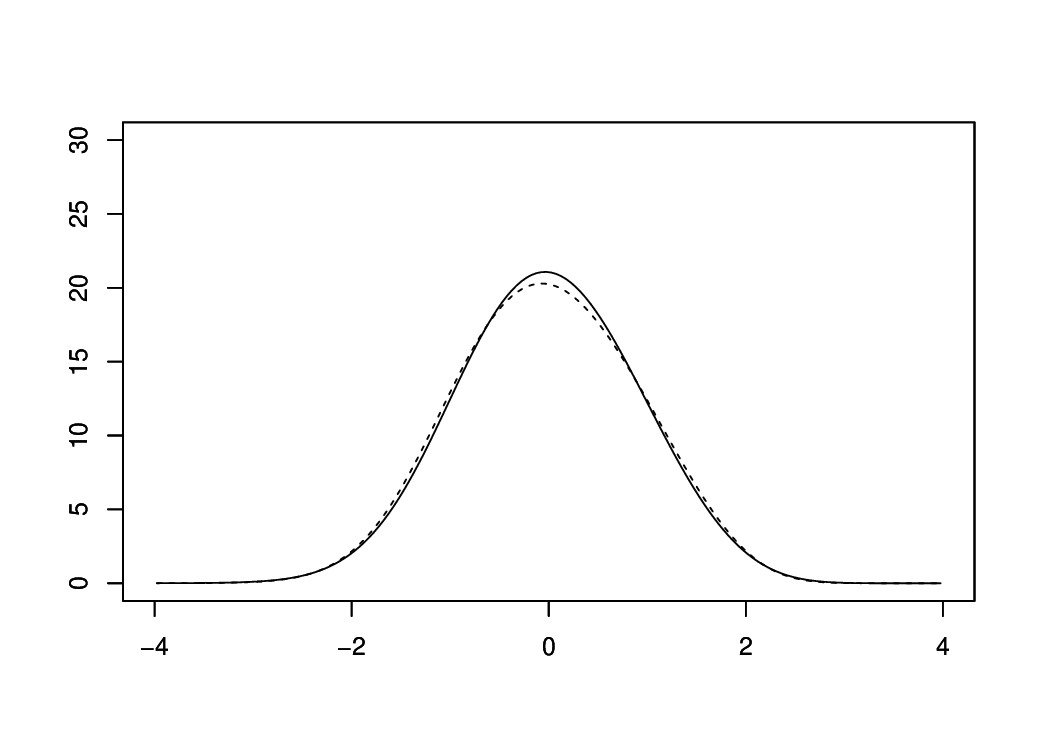}} \\
\caption{Distribution of posterior scores estimated by Poisson regression (all null groups). Straight lines represent results using robust standard deviations; dotted lines represent results using scores obtained by the inverse function} \vspace{0cm}
\end{center}
\label{fig:2}
\end{figure}

\noindent

\begin{table}[hbtp]
  \caption{Simulation data sets}
  \label{table:data_type}
  \centering
  \begin{tabular}{cll}
    \hline
    &null data  & non-null data   \\
    \hline 
	Simulation 1 \quad \quad \quad \quad& $\beta_j=0$  & $\beta_j=1$\\
	Simulation 2 \quad \quad \quad \quad& $\beta_j=0$  & $\beta_j=3$\\
	Simulation 3 \quad \quad \quad \quad& $\beta_j=0$  & $\beta_j \stackrel{D}{\sim} N(1,1)$\\
	Simulation 4 \quad \quad \quad \quad& $\beta_j=0$  & $\beta_j \stackrel{D}{\sim} N(3,1)$\\
	Simulation 5 \quad \quad \quad \quad& $\beta_j=0$  & $\beta_j \stackrel{D}{\sim} Exp(1)$\\
	Simulation 6 \quad \quad \quad \quad& $\beta_j=0$  & $\beta_j \stackrel{D}{\sim} Exp(\frac{1}{3})$\\
        \hline
  \end{tabular}
\end{table}

\begin{figure}[ht]
\begin{center}
\subfigure[non-null $\beta_j=1$]{
\includegraphics[width=50mm,height=50mm]{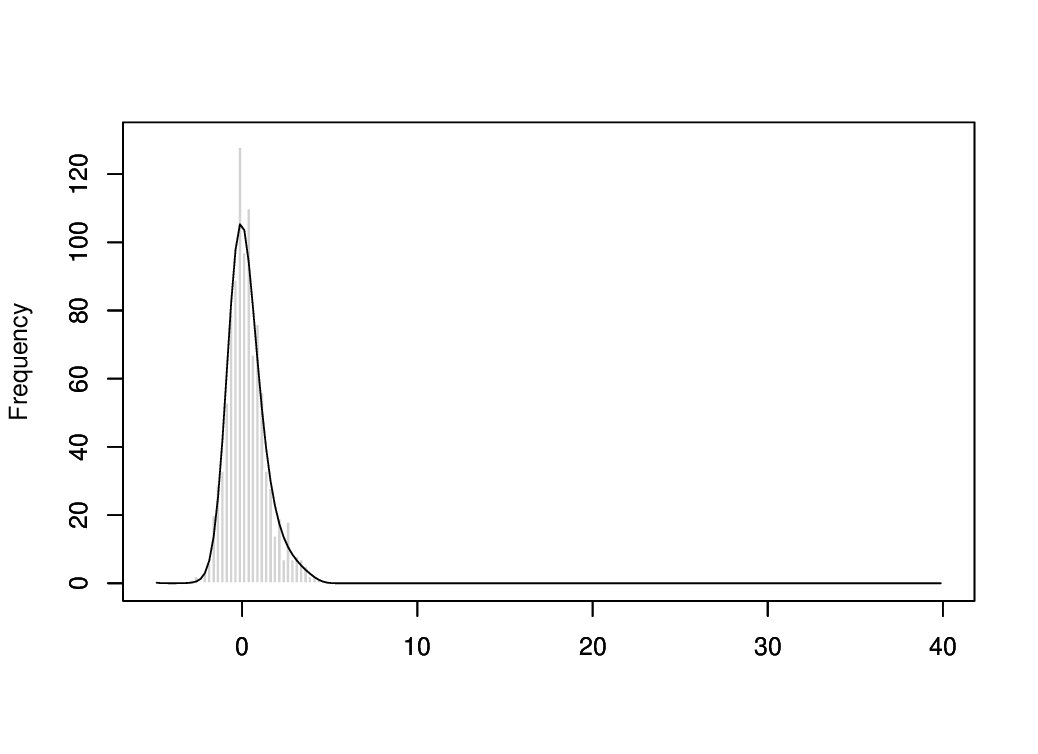}}
\subfigure[non-null $\beta_j=3$  ]{
\includegraphics[width=50mm,height=50mm]{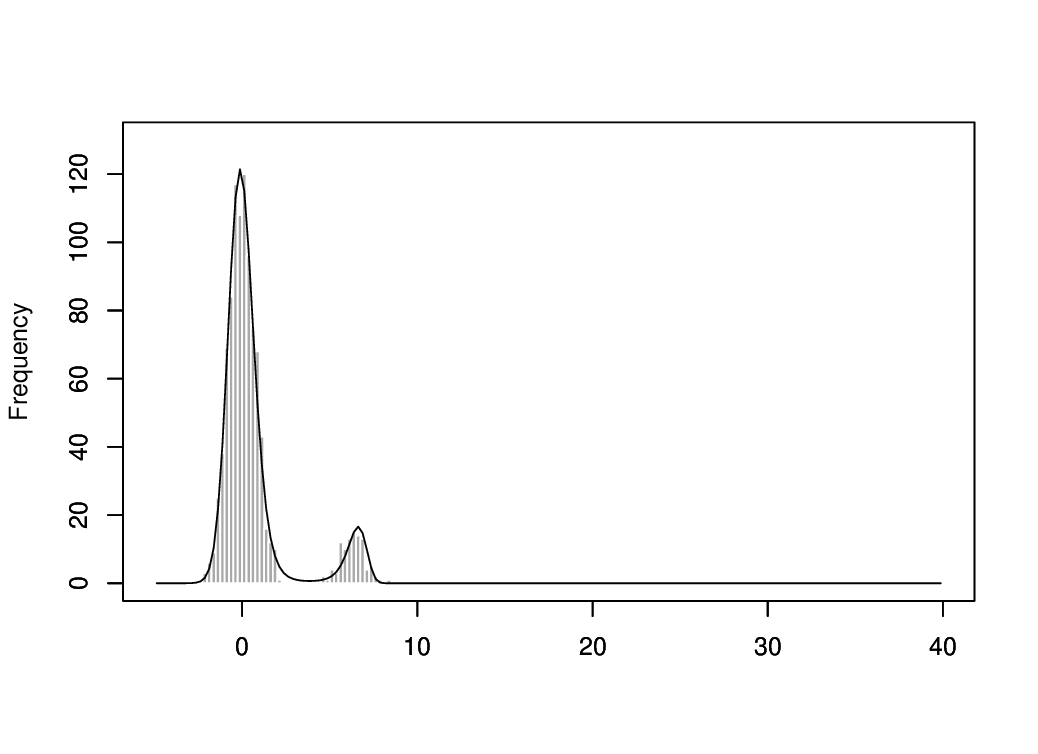}} \\
\subfigure[non-null $\beta_j \stackrel{D}{\sim} N(1,1)$]{
\includegraphics[width=50mm,height=50mm]{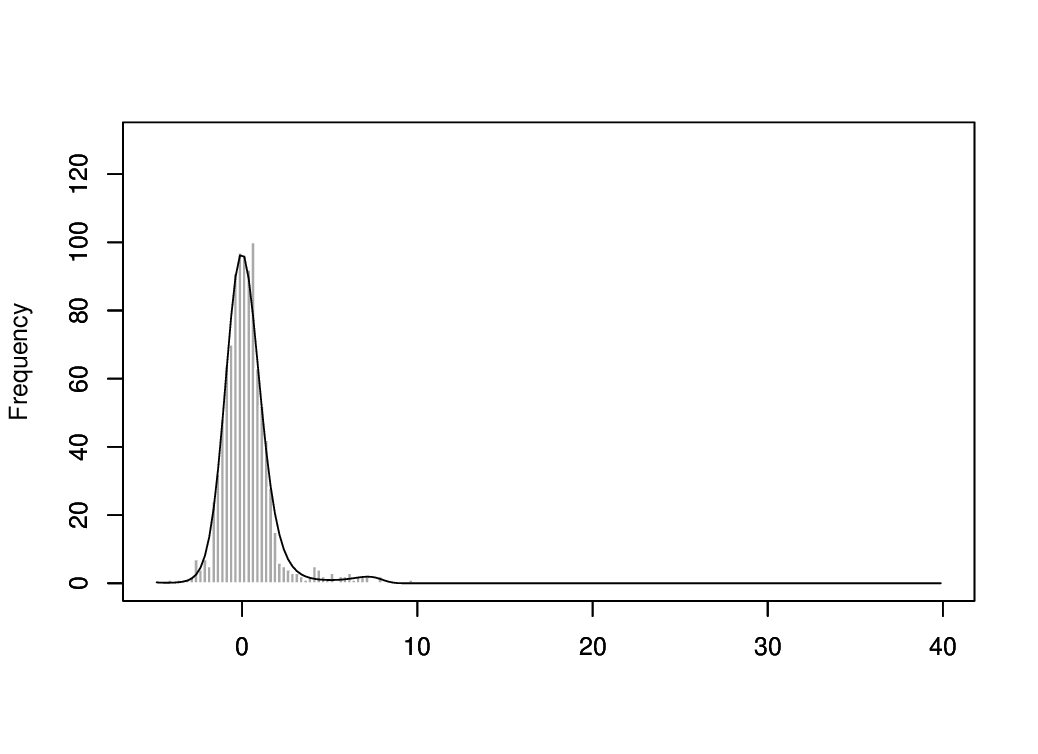}}
\subfigure[non-null $\beta_j \stackrel{D}{\sim} N(3,1)$ ]{
\includegraphics[width=50mm,height=50mm]{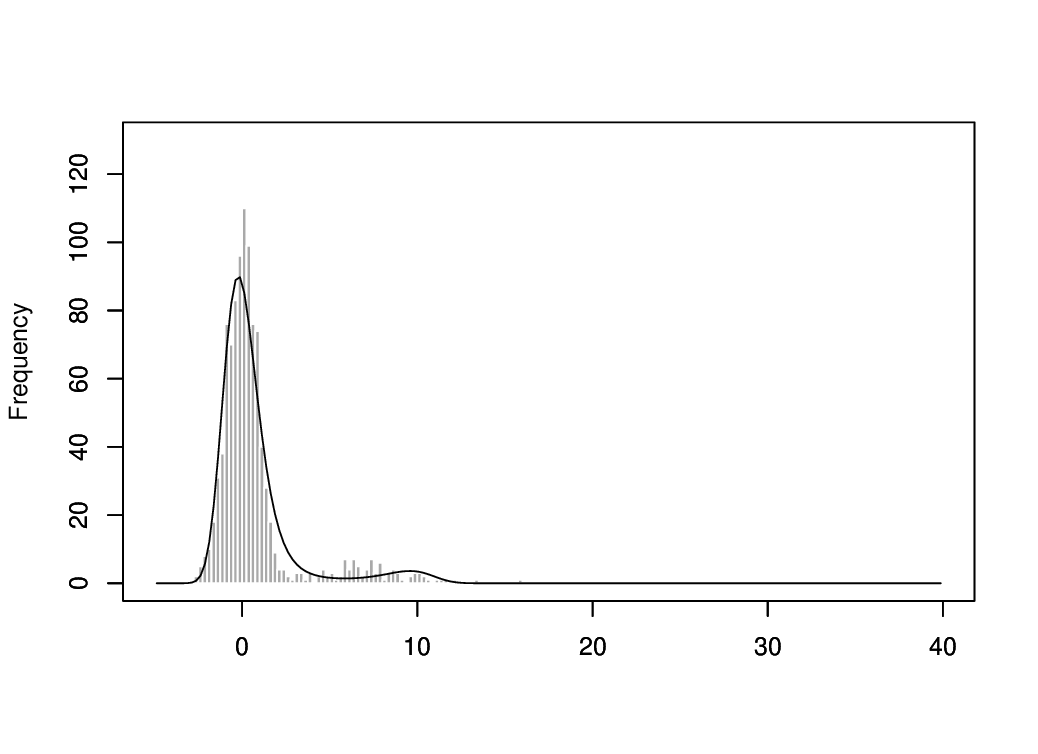}} \\
\subfigure[non-null $\beta_j \stackrel{D}{\sim} Expon(1)$]{
\includegraphics[width=50mm,height=50mm]{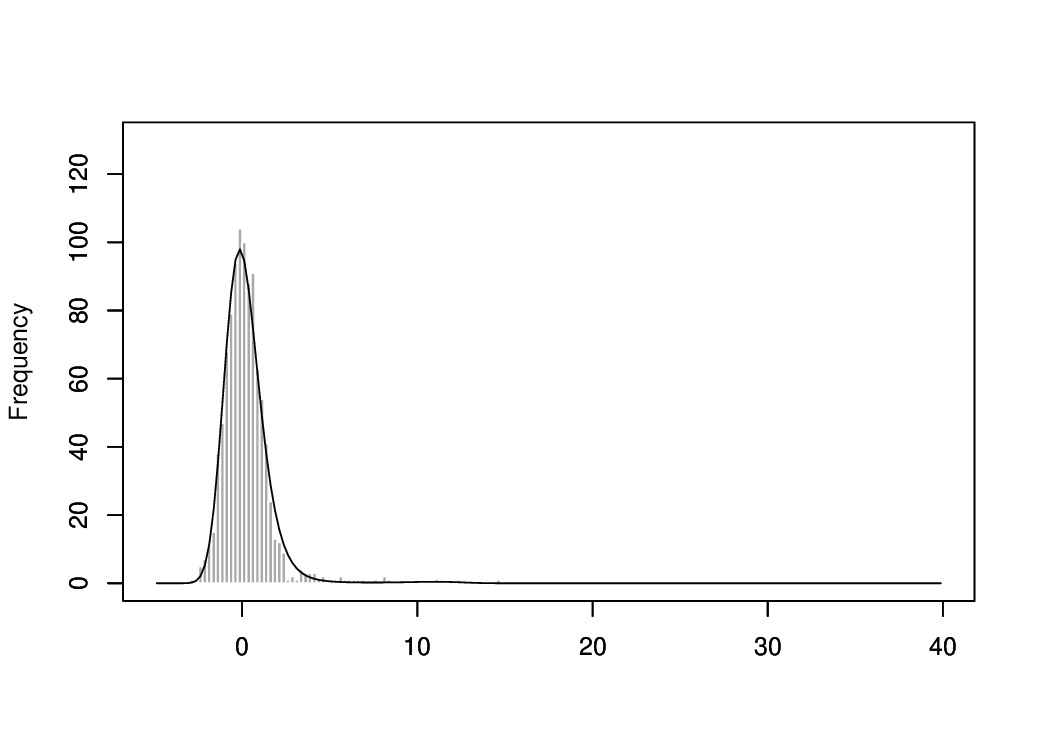}}
\subfigure[non-null $\beta_j \stackrel{D}{\sim} Expon(1/3)$ ]{
\includegraphics[width=50mm,height=50mm]{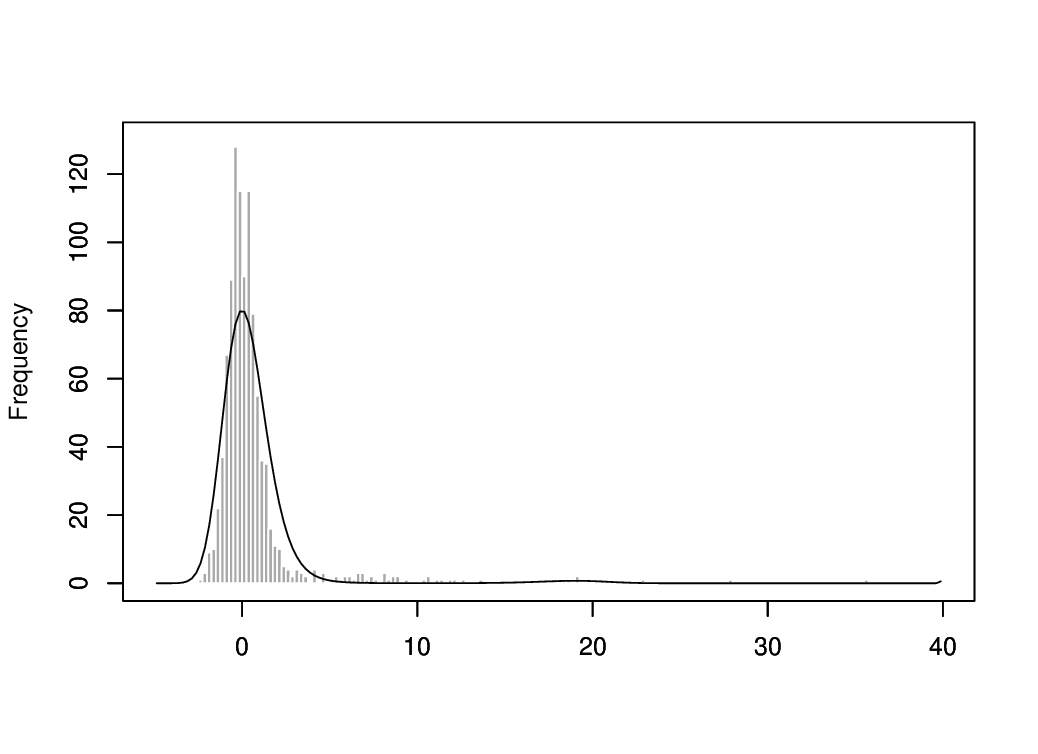}} \\
\caption{Histograms and estimated distributions of posterior scores} \vspace{0cm}
\end{center}
\label{fig:1}
\end{figure}

\begin{figure}[ht]
\begin{center}
\subfigure[non-null $\beta_j=1$]{
\includegraphics[width=50mm,height=50mm]{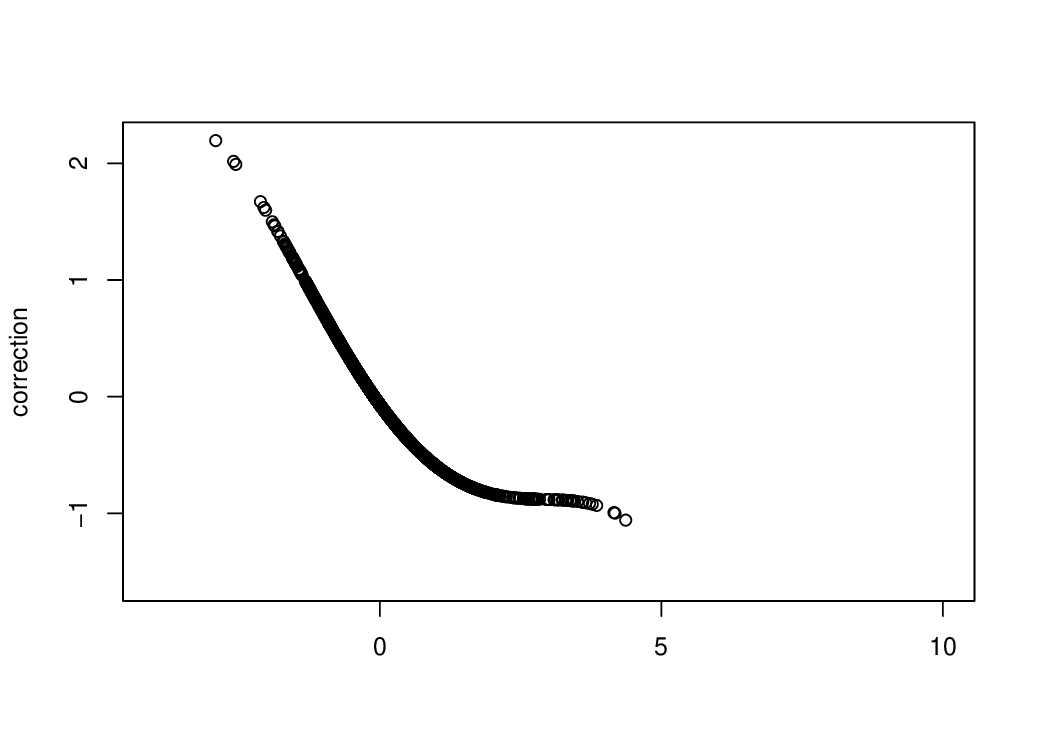}}
\subfigure[non-null $\beta_j=3$  ]{
\includegraphics[width=50mm,height=50mm]{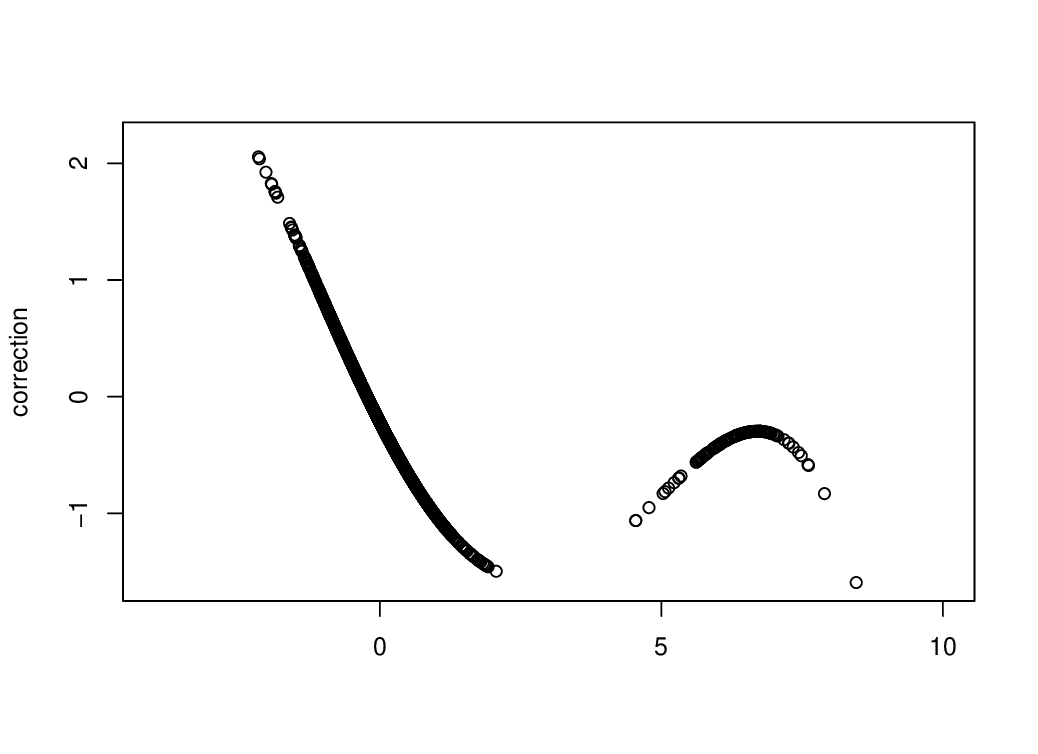}} \\
\subfigure[non-null $\beta_j \stackrel{D}{\sim} N(1,1)$]{
\includegraphics[width=50mm,height=50mm]{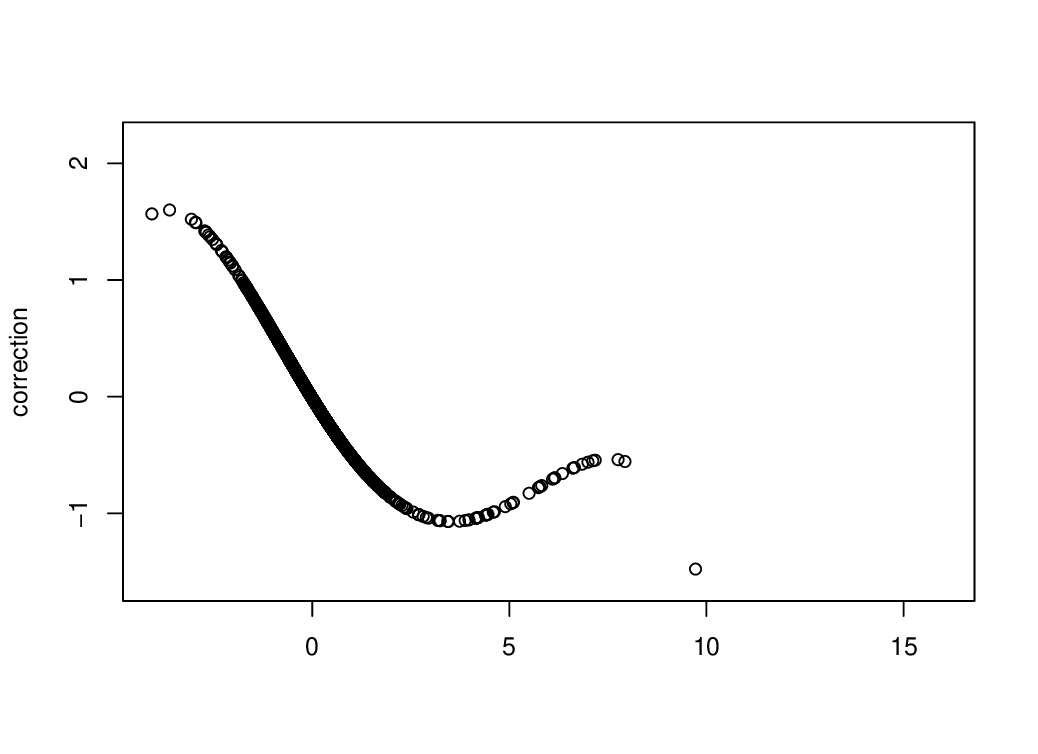}}
\subfigure[non-null $\beta_j \stackrel{D}{\sim} N(3,1)$ ]{
\includegraphics[width=50mm,height=50mm]{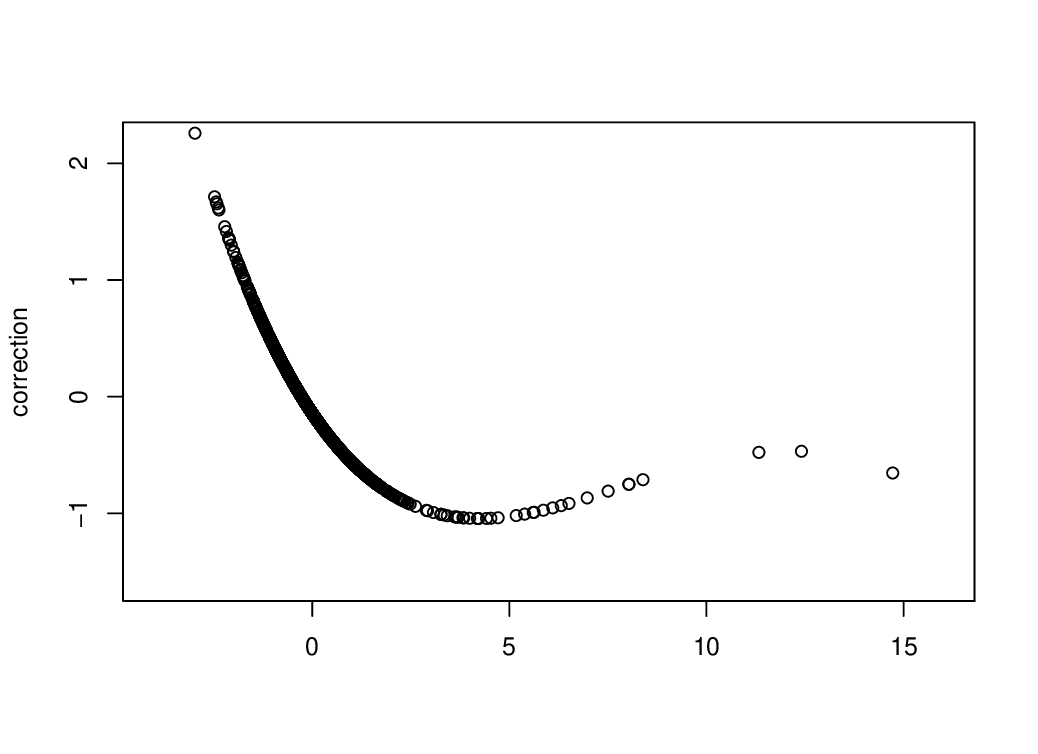}} \\
\subfigure[non-null $\beta_j \stackrel{D}{\sim} Expon(1)$]{
\includegraphics[width=50mm,height=50mm]{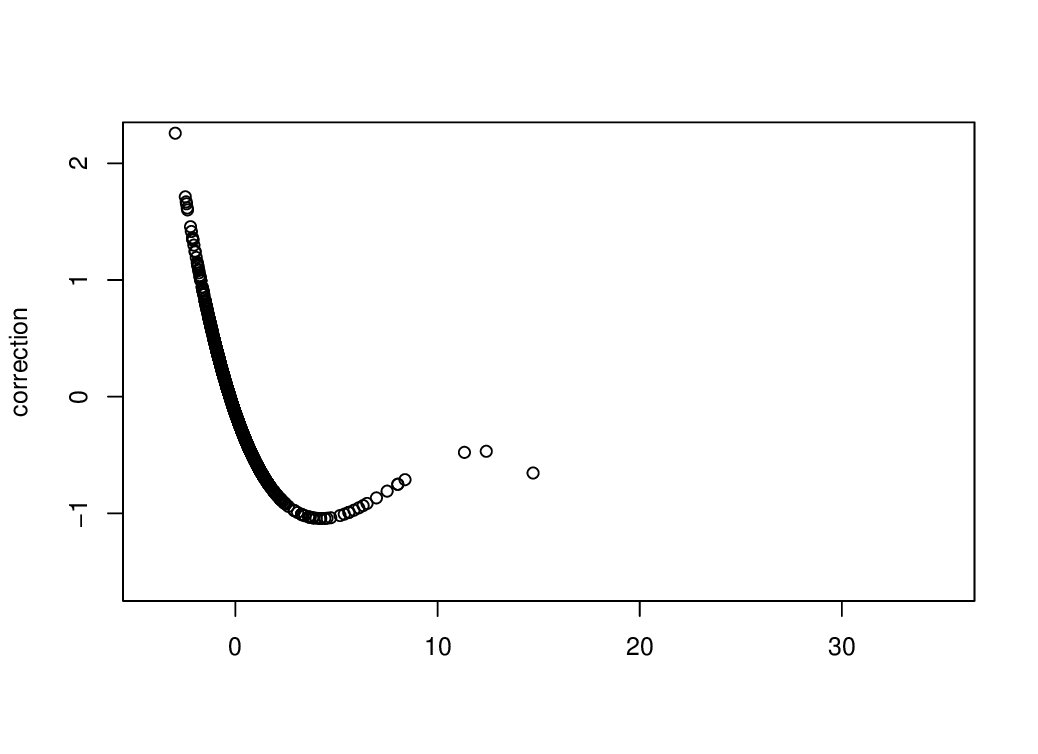}}
\subfigure[non-null $\beta_j \stackrel{D}{\sim} Expon(1/3)$ ]{
\includegraphics[width=50mm,height=50mm]{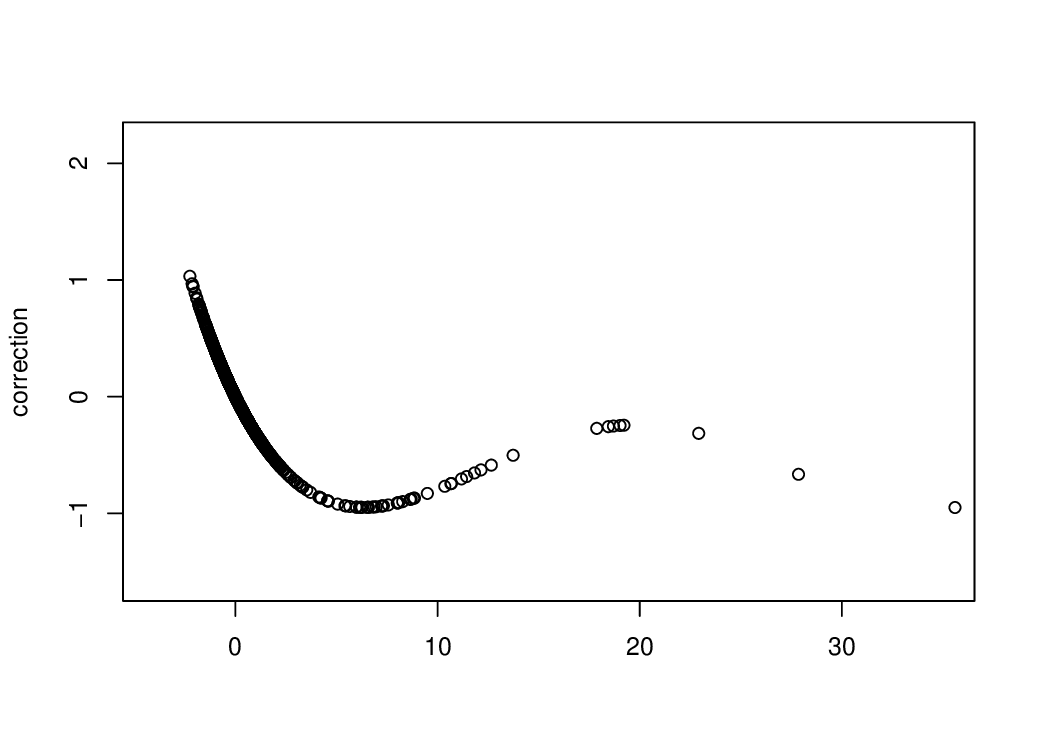}} \\
\caption{Modification term}\vspace{0cm}
\end{center}
\label{fig:4}
\end{figure}

\begin{figure}[ht]
\begin{center}
\subfigure[non-null $\beta_j=1$]{
\includegraphics[width=50mm,height=50mm]{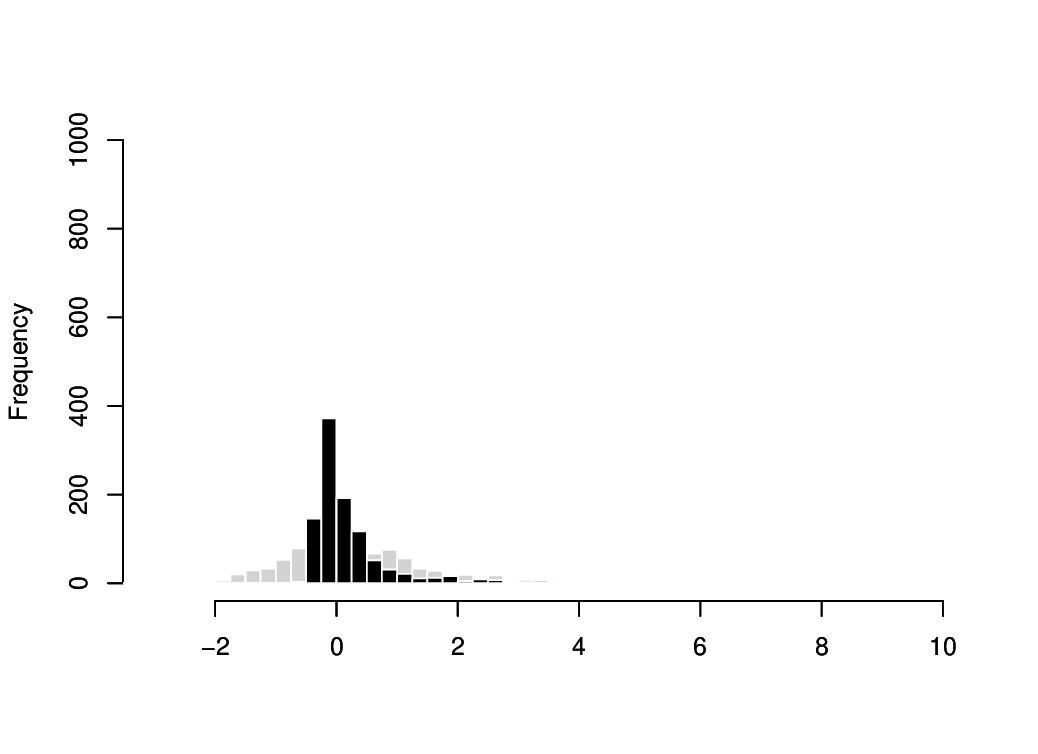}}
\subfigure[non-null $\beta_j=3$  ]{
\includegraphics[width=50mm,height=50mm]{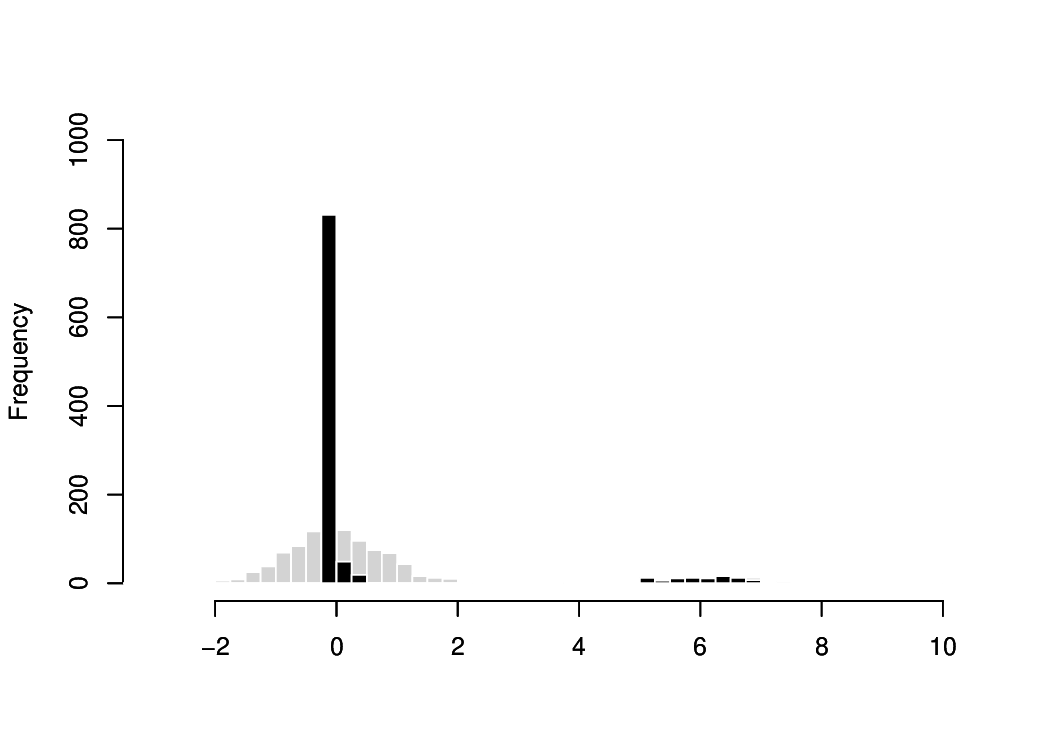}} \\
\subfigure[non-null $\beta_j \stackrel{D}{\sim} N(1,1)$]{
\includegraphics[width=50mm,height=50mm]{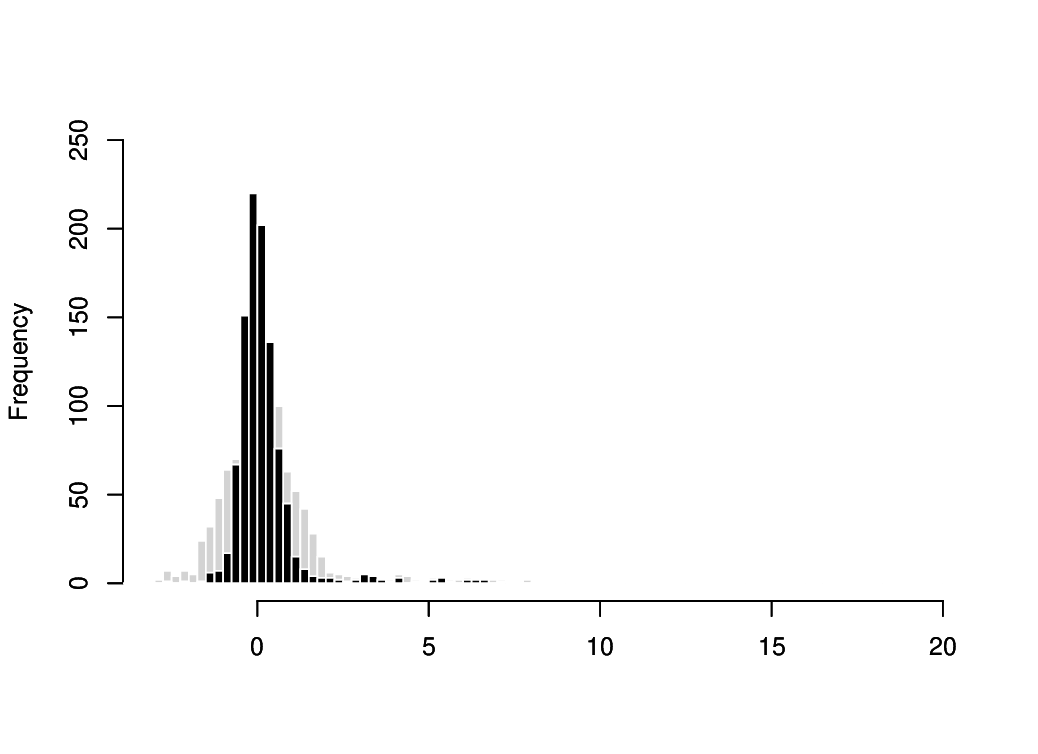}}
\subfigure[non-null $\beta_j \stackrel{D}{\sim} N(3,1)$ ]{
\includegraphics[width=50mm,height=50mm]{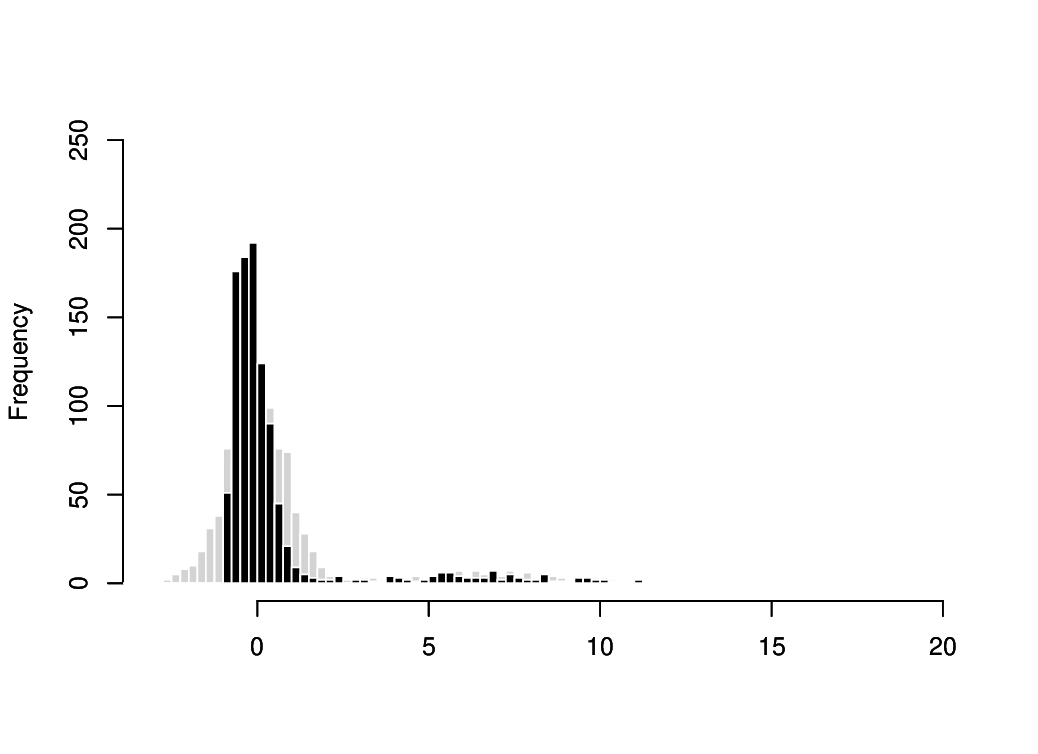}} \\
\subfigure[non-null $\beta_j \stackrel{D}{\sim} Expon(1)$]{
\includegraphics[width=50mm,height=50mm]{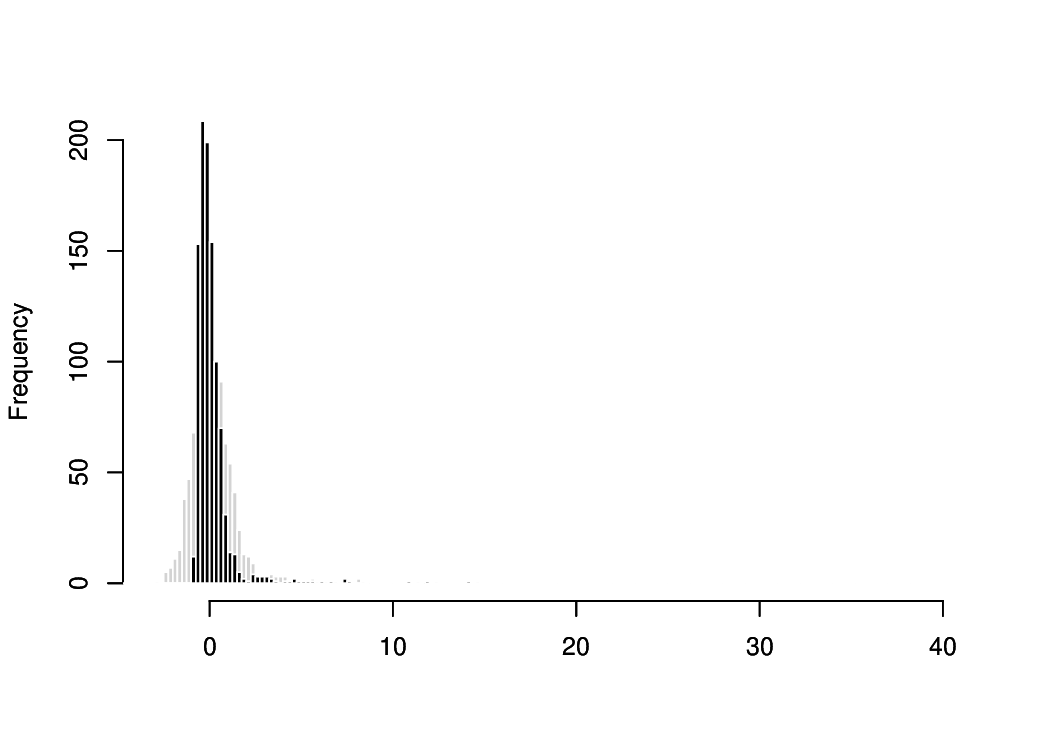}}
\subfigure[non-null $\beta_j \stackrel{D}{\sim} Expon(1/3)$ ]{
\includegraphics[width=50mm,height=50mm]{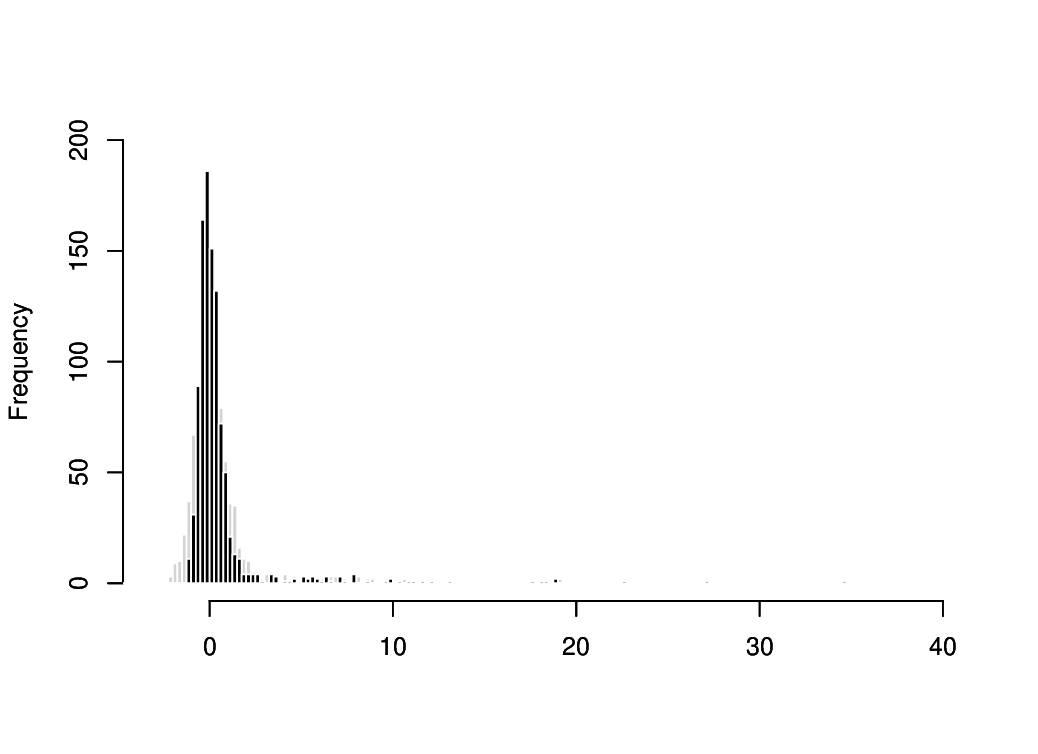}} \\
\caption{Histograms before (light gray) and after (black) correction}\vspace{0cm}
\end{center}
\label{fig:5}
\end{figure}

where the exponential distribution is defined by $f(x|\lambda)=\lambda \exp^{\lambda x}$ 
Figure 4 plots the Bayes correction terms. The  Bayes modification term takes a straght line when the model is approximated by a second-order spline. From these figures, it can be seen that in both scenarios, the modification is based on two modes. In other words, the hyper-prior distribution takes more than one mode. Figure 5 compares the results before and after the empirical Bayesian correction. The light gray histograms show the original posterior score, and the black histograms show the corrected scores. We can see that the correction works quite well.

\section{Conclusion}

The study of large-scale data has made significant progress in statistics over the past decades. In problems involving large-scale data, there are many target distributions, and they are mixed distributions with means that cannot be considered sufficiently close. In recent years, several studies have attempted to investigate the bias correction method involving such large-scale data. In this paper, we examine the applicability of the local empirical Bayes correction by Efron (2011) to Bayesian modeling for large-scale data.
When considering Bayesian modeling for large data sets, the prior distribution set by Efron (2011) plays the role of a hyperprior distribution. Since these priors have a significant impact on the results, we have to be careful. In this paper, we attempted to apply a local empirical Bayes correction based on the posterior predictive probabilities of the Bayesian model, without assuming a hyperprior distribution. We obtain good correction results.

\clearpage

\section{References}
\begin{enumerate}

\item Burkardt, J. (2014), The Truncated Normal Distribution, unpublished manuscrit

\url{http://people.sc.fsu.edu/jburkardt/presentations/truncated normal.pdf}

\item Carlin, B. P. and Louis, T. A. (2000), Bayes and Empirical bayes methods for data analysis 2nd ed., Chapman \& Hall.

\item Efron, B. (2011). Tweedie's Formula and Selection Bias. Journal of the American Statistical Association 106(496), 1602-1614.

\item Hawinkel, S. Thas, O. and Maere, S. (2023). The winner's curse under dependence: repairing empirical Bayes and a comparative study. 10.1101/2023.09.22.558978. 

\item Hogg, R. V. and Tanis, E. A. (2009), \textit{Probability and Statistical Inference}, Pearson.

\item Koenker, R. and Gu, J. (2016). REBAYES:  An R package for empirical Bayes mixture methods, \url{http://jiayinggu.weebly.com/uploads/3/8/9/3/38937991/rebayes.pdf}


\item Robbins, H. (1956) An empirical Bayes approach to statistics. In Proceedings of the Third Berkley Symposium on Mathematical Statistics and Probability, 1954-55, vol. 1 157-163.

\item Simon, N. and Simon, R. (2013). On estimating many means, selection bias, and the bootstrap. arXiv.1311.3709v1

\item Storey JD, Tibshirani R. (2003) Statistical significance for genomewide studies. Proceedings of the National Academy of Sciences. 100(16):9440-5.

\item Tan, K. M. , Simon, N. and Witten, D (2015). Selection bias correction and effect size estimation uncer dependence. arXiv.1405.4251v2

\item Wager S. (2014) A geometric approach to density estimation with additive noise, Statistica Sinca 24 533-554.


\item Ventrucci, Massimo and Scott,  Marian, E. (2011) "Multiple testing on standardized mortality ratios: a Bayesian hierarchical model for FDR estimation," Biostatistics, vol.12 no. 1 pp.55-67.
\end{enumerate}

\end{document}